\documentclass[12pt]{report}

\usepackage{graphicx}
\usepackage{amssymb}
\usepackage{multirow}
\usepackage{multicol}
\usepackage[square,numbers]{natbib}
\usepackage{longtable}
\usepackage{fancyhdr}
\usepackage{tikz}
\usepackage[tikz]{mdframed}

\usepackage{color, soul}

\usepackage{titlesec} 
\usepackage[left=4cm,right=4cm,top=2.5cm,bottom=4cm]{geometry}

\definecolor{DarkRed}{RGB}{163,22,31}
\definecolor{LightRed}{RGB}{247,14,45}
\definecolor{PaleRed}{RGB}{239,134,148}
\definecolor{cpiGray}{RGB}{106,100,100}

\titleformat*{\section}{\color{DarkRed}\normalfont\bfseries\LARGE}
\titleformat*{\subsection}{\color{LightRed}\normalfont\bfseries\LARGE}
\titleformat*{\subsubsection}{\color{PaleRed}\normalfont\bfseries\large}

\title{Characterising Cybercriminals: A Review}
\author{Matthew Edwards, Emma Williams, Claudia Peersman \& Awais Rashid}

\renewcommand{\maketitle}{\newpage
\newgeometry{margin = 0in}
\includegraphics[width=3.09in]{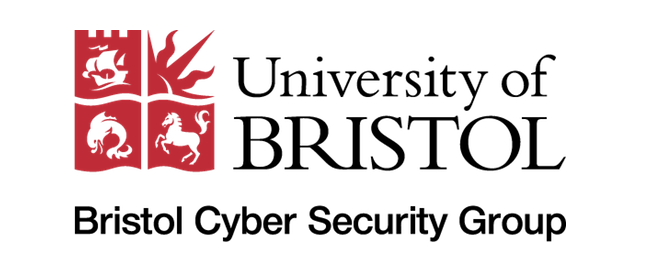}
\setlength{\fboxsep}{0pt}
\hfill \colorbox{cpiGray}{\makebox[3.22in][r]{\shortstack[r]{\vspace{2.75in}}}}%
\vspace{-0.25pt}
\setlength{\fboxsep}{10pt}
\setlength{\fboxrule}{0pt}
\colorbox{DarkRed}{\makebox[8.25in][l]{\hfill \shortstack[r]{\fontsize{36}{36}\rmfamily\color{white} Characterising Cybercriminals: A Review\\%
\fontsize{24}{24}\rmfamily\color{white}}}}%
\setlength{\fboxsep}{0pt}
\vspace{-8.5pt}
\hfill \colorbox{cpiGray}{\hspace{.25in} \parbox{2.97in}{\vspace{4in} \color{white} \textbf{Matthew Edwards \\ Emma Williams \\ Claudia Peersman \\ Awais Rashid \\\\   \today \vspace{2.3in} \vfill}}}%
\let\Title\title
\restoregeometry}

\mdfdefinestyle{SummaryBox}{%
    linecolor=DarkRed,
    outerlinewidth=3pt,
    roundcorner=20pt,
    innertopmargin=\baselineskip,
    innerbottommargin=\baselineskip,
    innerrightmargin=20pt,
    innerleftmargin=20pt,
    frametitlerule=true,
    frametitlerulecolor=DarkRed,
    frametitlefontcolor=white,
    frametitlebackgroundcolor=DarkRed,
    fontcolor=DarkRed,
    frametitle=Summary,
    backgroundcolor=gray!50!white,
    leftmargin=18pt,
    rightmargin=18pt}
    
\mdfdefinestyle{CharBox}{%
    linecolor=DarkRed,
    outerlinewidth=3pt,
    roundcorner=20pt,
    innertopmargin=\baselineskip,
    innerbottommargin=\baselineskip,
    innerrightmargin=20pt,
    innerleftmargin=20pt,
    frametitlerule=true,
    frametitlerulecolor=DarkRed,
    frametitlefontcolor=white,
    frametitlebackgroundcolor=DarkRed,
    fontcolor=DarkRed,
    frametitle=Characteristics,
    backgroundcolor=gray!50!white,
    leftmargin=18pt,
    rightmargin=18pt}

\mdfdefinestyle{MotivBox}{%
    linecolor=DarkRed,
    outerlinewidth=3pt,
    roundcorner=20pt,
    innertopmargin=\baselineskip,
    innerbottommargin=\baselineskip,
    innerrightmargin=20pt,
    innerleftmargin=20pt,
    frametitlerule=true,
    frametitlerulecolor=DarkRed,
    frametitlefontcolor=white,
    frametitlebackgroundcolor=DarkRed,
    fontcolor=DarkRed,
    frametitle=Motivations,
    backgroundcolor=gray!50!white,
    leftmargin=18pt,
    rightmargin=18pt}

\begin{document}

\begin{titlepage}
\maketitle
\end{titlepage}










\section*{Executive Summary}

This review provides an overview of current research on the known \emph{characteristics} and \emph{motivations} of offenders engaging in cyber-dependent crimes. Due to the shifting dynamics of cybercriminal behaviour, and the availability of prior reviews in 2013, this review focuses on original research conducted from 2012 onwards, although some older studies that were not included in prior reviews are also considered. As a basis for interpretation of results, a limited \emph{quality assessment} was also carried out on included studies through examination of key indicators.

The results of this survey indicate that while the research base is expanding, evidence on these topics remains limited. Existing research is constrained by methodological limitations and difficulties in accessing offender populations. Many studies reported on low-level or loosely-defined forms of cybercriminality, self-reported in student or youth populations, rather than technically sophisticated attackers. Anonymous online surveys of cybercriminal venues have greater potential to access this latter population, but are limited by doubt about the accuracy of self-reported accomplishments or skill. Police case data is also used, but relatively small numbers of prosecutions for cybercrime relative to traditional offending hamper generalisability of results, particularly when focused on technically sophisticated attackers.

With these limitations in mind, the characteristics supported from the literature include a greater likelihood of being in younger age groups, being male, high levels of internet/computer use, completing or having completed higher education, and low levels of self-control. In Western countries, there may be an ethnic bias towards offenders being white. Limited evidence exists on the living situations of offenders, with some indication that being single is more likely amongst cybercriminals. 

The motivations highlighted from the literature include financial reward, revenge, previous victimisation in cybercrime, a desire for power and recognition, curiosity, immersion in the challenge of defeating security systems, the perception of low risk of punishment, and perceived social and moral acceptability of cybercrime. Revenge and previous victimisation are given particular attention, with examination of the overlap between offenders and victims being a focus of a number of studies. 

The review results highlight the need for more objective measures of cybercriminals, and greater focus on technically sophisticated attacker populations. 
The AMoC project's later stages will tackle these challenges through large-scale data mining of cybercriminal forums.

\pagebreak

\section{Introduction}
\label{sec:intro}
This literature review aims to provide an overview of current research regarding the characteristics and motivations of offenders engaging in cyber-dependent crimes. It represents the first deliverable within WP1 of the AMoC project and reviews existing literature on cyber offenders, their motivations and characteristics (both in terms of demographic characteristics and psychological characteristics). This review will form the basis for the development of an initial framework regarding cybercriminal characteristics,  providing an evidence-based foundation from which to interpret data gathered from forums and other sources, both over the course of the AMoC project and in future research. 

The review is structured as follows: 
\begin{itemize}
\item The remainder of Section~\ref{sec:intro} considers (i) the method used to conduct the literature review, (ii) the primary theoretical approaches that have been applied to aid understanding of the behaviour, characteristics and motivations of cyber offenders, and (iii) a summary of work conducted prior to the main period of focus for the literature search.
\item Section~\ref{sec:characteristics} provides an overview of the primary characteristics of cyber offenders that have been identified within prior literature and considers the relative robustness of findings in this area, focusing predominantly on literature published from 2012 onward. 
\item Section~\ref{sec:motivations} discusses  the various motivations of cyber offenders that have been identified within prior literature, again considering the relative robustness of findings in this area and focusing predominantly on literature published from 2012 onward.
\item Section~\ref{sec:quality} reports on the overall quality assessment of literature included in the review, focusing on sample sizes and types as well as method replicability.
\item Section~\ref{sec:limitations} discusses limitations of existing research in this area, including the primary research methods used and associated issues. This section also highlights current gaps within the existing research base, which may limit the extent to which robust conclusions can be drawn.
\item Finally, Section~\ref{sec:conclusions} offers key conclusions from the literature review, including the main characteristics and motivations that are sufficiently supported to warrant inclusion within the initial framework.
\end{itemize}
   
\subsection{Review Methodology}
This review focuses on research that investigates the various \textit{motivations} of cyber offenders (i.e., the factors that account for \textit{why} an individual engages in online offending) and the likely \textit{characteristics} of these offenders, both in terms of demographic characteristics (i.e., age, gender, education level, etc.) and psychological characteristics (e.g., personality traits, behavioural characteristics). Research that does not investigate these aspects will be excluded from the review. This review will also focus on literature investigating particular types of cyber offences, namely \textit{cyber-dependent crimes} such as hacking, malware development, online fraud (e.g., carding), and hacktivism. Research investigating cyber terrorism, child sexual exploitation or other forms of cyber-enabled crime will be explicitly excluded. 

The review will focus on identifying and evaluating \textit{original research} conducted in this area, although review articles and those describing or evaluating relevant theoretical models may also be considered if they make an original contribution to the research base. Existing research will also be subject to a form of \textit{quality assessment} to enable the potential contribution, generalisability and likely replicability of the research to be evaluated. This quality assessment will focus on considering the participant sample size and sample type (e.g., the relative diversity of participants), the publishing outlet (e.g., the impact factor of academic journals, whether the article has undergone peer-review prior to publication), and whether sufficient information is provided within the research methodology of the article to enable the method to be effectively reproduced by other researchers if they had access to a comparable sample population. The extent to which particular findings have been replicated in multiple research studies will also be considered. 

The fast pace of change inherent in online services, digital technologies, and resultant criminal activities exploiting these services is likely to result in an ever-shifting dynamic of cybercriminal behaviours, motivations and characteristics. To ensure that this literature review considers an evidence-base that is sufficiently relevant to current issues within the time available, the literature search will focus predominantly on identifying and evaluating research that has been published from \textit{2012 onward}. Research published prior to this time will not be ignored, but will similarly not form the predominant basis for evidence-based recommendations and evaluations. This date was specifically chosen due to the publication of a number of in-depth reviews of research regarding the characteristics and motivations of various types of cyber offenders in 2013 (e.g., \citet{kirwan2013cybercrime,broadhurst2013crime}), and a Home Office review of available evidence on cybercrime more generally~\cite{mcguire2013cyber}, which includes a focus on UK evidence about offender characteristics.  Together, these resources provided a strong basis for understanding the key findings of research prior to this period.

\subsection{Key Theoretical Approaches}
A range of theoretical approaches from psychology, criminology and sociology have previously been considered in relation to the behaviour, motivations and characteristics of cyber criminals. These have focused predominantly on five main theories: (1) neutralisation theory, (2) self-control theory, (3) social cognitive theory, (4) routine activity theory, and (5) the theory of planned behaviour.

\subsubsection{Neutralisation Theory}
Neutralisation Theory~\citep{sykes1957techniques} focuses on how individuals perceive and consider the potential impact of their criminal activities, and the way in which they may distort these perceptions in order to make their activities more morally acceptable. This can be linked to the psychological process of cognitive dissonance~\citep{festinger1957theory}, whereby people find it uncomfortable to hold competing thoughts about an issue. For instance, if they believe that engaging in hacking is a morally unethical behaviour, but they engage in this activity, then in order to reduce the resultant conflict between their moral perceptions and their behaviours, they can either stop engaging in the behaviour altogether (i.e., no longer engage in hacking activities), or they can adjust their attitudes and thoughts to correspond with their behaviour (e.g., considering that hacking is not hurting anyone directly and so it must be okay, thus making it a morally acceptable activity). Neutralisation theory has been applied to hacking (e.g., \cite{morris2011computer}) and digital piracy activities (e.g., \cite{morris2009neutralizing}), with 5 primary mechanisms originally highlighted by which this may occur. These include: (a) denial of responsibility (e.g., ``I had no other choice''), (b) denial of injury (e.g., ``it doesn't hurt anyone''), (c) denial of victim (e.g., ``the victim deserves it in some way''), (d) condemnation of the condemners (e.g., ``victims are hypocrites'') and (e) appeal to higher loyalties (e.g., the behaviour was warranted to achieve a higher purpose). 

\subsubsection{Self-Control Theory}
The Self-Control Theory of Crime considers that individuals often engage in criminal activities due to failure in their ability to effectively control their behaviour~\cite{muraven2006self}.  Self-control is considered a trait psychological characteristic linked to the ability to avoid and resist temptation (or not) (see \citet{ent2015trait}). Individuals low in self-control are considered to be more impulsive, take greater risks, and fail to adequately consider the long-term consequences of their actions. This trait has also been linked with cyber criminal activities, such as downloading illegal content~\cite{higgins2007digital}, as well as traditional forms of crime.

\subsubsection{Social Cognitive Theory}
Social Cognitive Theory~\cite{bandura1986social} reflects how individuals learn how to behave in a particular environment based on observing the actions of those around them and the resultant consequences and outcomes of the behaviours that they observe (both positive and negative). It is proposed that this social learning process occurs through engagement with particular role models within an individual's environment, such as observing the behaviour of family members. It can also occur via exposure to mass-media content or through observation and understanding of particular social norms and activities linked to specific geographical locations or communities. This process may not only influence what types of behaviour an individual considers to be socially acceptable, but can also lead to direct learning and skill acquisition related to the actual ability to engage in particular types of deviant online activities. For instance, attitudes to digital piracy have previously been found to be related to association with deviant peers and the perception that important others wanted or expected the individual to engage in digital piracy behaviours~\citep{kirwan2013cybercrime,d2005music,malin2009adolescent}. Therefore, engagement and affiliation with others who engage in cyber criminal behaviours can influence an individual's perceptions of the acceptability of such activities, social norms and expectations surrounding these behaviours, and the actual ability to effectively engage in cyber criminal activities. Similarly, social-ecological approaches to crime and deviant online activities, such as cyber bullying, highlight the potential role of these various social factors in influencing individual behaviour (e.g., close peer and family relationships, wider community settings, and societal-level norms and approaches)~\citep{cross2015social}. 
 
\subsubsection{Routine Activities Theory}
Routine Activities Theory~\cite{cohen1979social} refers to three main elements that must align in order for a particular crime to occur: (1) the existence of an attractive target (e.g., credit card details stored in a particular location), (2) the presence of a motivated offender (e.g., an individual who becomes aware of this data or actively seeks it out) and (3) the lack of a capable guardian (e.g., a lack of sufficient technical protections). As a result, the existence of a sufficiently attractive target and the use of poor protective mechanisms can, in turn, influence the behaviour, characteristics and potential motivations of an offender to commit a particular crime (e.g., a particular activity representing a sufficient intellectual challenge, the lack of a requirement for technical skills to achieve a particular outcome or use a particular method, the potential availability of different types of targets that relate to various degrees of financial or reputational motivations).

\subsubsection{The Theory of Planned Behaviour}
The Theory of Planned Behaviour (TPB)~\cite{ajzen1991theory} provides a framework to explore the relationship between an individual's beliefs and their behaviour. It considers three primary mechanisms through which an individual's intentions to engage in a particular behaviour may be influenced: (1) their attitudes towards a particular behaviour, which represents how positively or negatively they evaluate the behaviour and those who engage in it, (2) perceived social norms related to engaging in that behaviour, and (3) the perceived ease of engaging in the behaviour (termed perceived behavioural control). Overall, the use of the TPB as a framework to understand and change behaviour has been supported in multiple domains, although effects are generally found to be stronger for intentions to engage in particular activities compared to actual behavioural responses (see \citet{sheeran2016intention} for a review). Although the TPB has been predominantly considered in relation to what factors influence engagement in secure online behaviour, there has also been a very limited consideration of how these concepts may also apply to cyber criminal behaviours, such as hacking (e.g., \citet{rennie2007advanced}). 
\subsubsection{Other Approaches}
Finally, previous research has highlighted that the likelihood of getting caught when engaging in a cyber criminal activity influences an individual's attitudes and resultant behaviour. For instance, \citet{nandedkar2012won} assessed the presence of optimism bias in relation to music piracy and suggested that greater optimism with regards to potential risks (i.e., thinking that they were less likely to be caught compared to other people) influenced attitudes towards digital piracy. Theoretical models originally applied to the health behaviour domain, such as Protection Motivation Theory~\cite{rogers1975protection}, have also shown how differences in perceived risk (i.e., the perceived likelihood of a threatening outcome occurring, and the potential impact of that threat if it were to occur) combine with other aspects,  such as the perceived ability to cope with the threat and the perceived benefits of a threat-reducing behaviour, to influence resultant protective actions, (e.g., engaging in secure online behaviour~\cite{tsai2016understanding}). However, this approach has yet to be applied to decisions to engage in cyber criminal behaviour and has particular relevance when considering potential responses to interventions and adverse events.

\subsection{Research Pre-2012}
A considered summary of research conducted prior to 2013 is provided by
\citet{kirwan2013cybercrime} , who discussed the state of research related to the motivations and characteristics of various types of cybercriminals in the 2013 book, Cybercrime: The Psychology of Online Offenders. This identified differences in the extent of research conducted across different types of cybercrime, with substantially more research focused on understanding the motivations and characteristics of hackers than on malware developers for instance, with research in the latter area being outdated and limited in its methodological approach. Conversely, digital piracy, which was considered to be a more widespread activity at the time, was found to be an area that demonstrated a more developed research base due to the relative ease of conducting research related to this activity with larger participant samples. 

Within their work, the authors highlighted how previous theoretical discussions related to hacking behaviour have emphasised the role of factors such as curiosity, the desire for achieving status in a social group, and the need for power, with more limited empirical research highlighting the role of practicing and developing skills, wishing to undertake simple pranks, and the desire for intellectual challenge. It was suggested that many hackers were younger males, who lived with parents, had achieved middle or higher levels of education, and who were often students or trainees using computers in their spare time. Limited research was also discussed that tentatively suggested that hackers displayed more nocturnal activities, were more task-oriented, had a higher propensity for risk, and demonstrated greater use of rational thinking styles rather than intuitive thinking styles. However, research in this area was found to lack consistent findings, as well as suffer from methodological issues related to the use of self-report survey methods, and the inherent constraints and biases associated with these. 

When considering other cyber offences, such as developing malware, \citet{kirwan2013cybercrime}  claimed that research suggested that such offenders were increasingly motivated by financial gain (although factors such as intellectual challenge, peer recognition, and vengeance were still present). Conversely, saving money and gaining immediate access to content were found to be key motivators for engaging in digital piracy behaviours, the latter of which can also be linked to psychological characteristics such as low self-control. Demographic characteristics associated with digital piracy were also found to include being younger and being more familiar with computers. 

Overall, the work of  \citet{kirwan2013cybercrime} highlights a range of different characteristics and motivations related to cyber criminal activities that were identified by early research, and that these were found to vary, both across different cyber offences and across different individuals engaging in the same type of offence, making the development of conclusive typologies challenging. The remainder of this review now considers how research in this area has advanced post-2012 to allow a more in-depth understanding of cybercriminal characteristics and motivations related to cyber-dependent crime.

\newpage
\section{Characteristics of Cybercriminals}
\label{sec:characteristics}

\begin{mdframed}[style=SummaryBox]
\begin{itemize}
	\item Studies were largely self-report surveys, often targeted at student or youth groups. 
	\item Under- and over-confidence in anonymous self-reporting may bias findings.
	\item Various measures of engagement with technology or the internet were found to be associated with cybercriminality.
	\item Most studies described cybercriminals as 60-90\% male, highly-educated and young.
	\item Widely studied demographic characteristics are summarised in Table~\ref{tab:characteristics}.
\end{itemize}
\end{mdframed}

\begin{table}
\footnotesize
\centering
\caption{Literature estimates of the common descriptive demographic characteristics of cybercriminals. The plurality response is given for most categorical
variables. Education was translated into primary or below, secondary, or tertiary and higher. }
\begin{tabular}{lrrr}
\hline
Characteristic & Estimate & Sample & Study \\
\hline
\multirow{11}{*}{Age} 	
			& ``late teens or 20s'' & 821 & \citet{hyslip2018defining} \\
		  	& 50\%+ $< 19$, 83\% $< 25$  & 729 & \citet{woo2003hacker} \\
			& 28.67 & 595 & \citet{liao2017computer} \\
			& 23 & 457 & \citet{voiskounsky2003flow} \\
			& 94\% $< 23$ & 179 & \citet{seigfried2015assessing} \\
			& 35.6 (m) 51.5 (f) & 119 & \citet{payne2018using} \\
			& 50.9\% 19-24, 33\% 25-30 & 112 & \citet{mcbrayer2014exploiting} \\
			& 30 & 61 & \citet{chan2014exploratory} \\
			& 29 & 39 & \citet{odinot2016cybercrime} \\
			& 34.71 & 14 & \citet{steinmetz2015becoming} \\
			& 90\% 16-24, 10\% 25-34 & 10 & \citet{hutchings2016exploring} \\
\hline
\multirow{9}{*}{Gender} & 88\% male & 821 & \citet{hyslip2018defining} \\
			& 88\% male & 729 & \citet{woo2003hacker} \\
			& 60.34\% male & 595 & \citet{liao2017computer} \\
 			& 49.2\% male & 179 & \citet{seigfried2015assessing} \\
			& 91.4\% male & 117 & \citet{payne2018using} \\
			& 51.4\% male & 112 & \citet{mcbrayer2014exploiting} \\
			& 67.2\% male & 61 & \citet{chan2014exploratory} \\
			& 92.9\% male & 14 & \citet{steinmetz2015becoming} \\
			& 80\% male & 10 & \citet{hutchings2016exploring} \\
\hline
\multirow{6}{*}{Ethnicity} 	& 63\% white & 821 & \citet{hyslip2018defining} \\
				& 60.4\% asian & 729 & \citet{woo2003hacker} \\			
				& 77.98\% white & 595 & \citet{liao2017computer} \\
				& 84.4\% white & 179 & \citet{seigfried2015assessing} \\
				& 89.3\% white & 112 & \citet{mcbrayer2014exploiting} \\
				& 92.9\% white & 14 & \citet{steinmetz2015becoming} \\
\hline
\multirow{4}{*}{Education} 	& ``the majority`` tertiary & 821 & \citet{hyslip2018defining} \\
				& 47.8\% tertiary & 729 & \citet{woo2003hacker} \\
				& 63.9\%  secondary & 61 & \citet{chan2014exploratory} \\ 
				& 100\% tertiary & 14 & \citet{steinmetz2015becoming} \\
				& 80\%+ tertiary & 10 & \citet{hutchings2016exploring} \\
\hline
\multirow{2}{*}{Marital Status}  & 68.8\% single & 112 & \citet{mcbrayer2014exploiting} \\
				& 63.9\% single & 61 & \citet{chan2014exploratory} \\
				& 50\% single & 14 & \citet{steinmetz2015becoming} \\
\hline
\multirow{2}{*}{Occupation}	& 55.7\% employed & 61 & \citet{chan2014exploratory} \\
				& 92.9\% employed & 14 & \citet{steinmetz2015becoming} \\
				& 80\% student & 10 & \citet{hutchings2016exploring} \\

\hline
\multirow{2}{*}{Criminal History}	& 90.2\% none & 90 & \citet{chan2014exploratory} \\
					& 42.1\% none & 19 & \citet{odinot2016cybercrime} \\

\hline
\end{tabular}
\label{tab:characteristics}
\end{table}

The role of a number of demographic and psychological factors in effectively characterising individuals engaged in cyber-dependent crime has been investigated in recent years, including age, gender, Internet usage, degree of self-control, and particular characteristics associated with Asperger's Syndrome, such as attention to detail and poor social skills. This section reviews research investigating these characteristics, with Table~\ref{tab:characteristics} providing a summary of findings related to demographic characteristics in this area.

In 2018, \citet{hyslip2018defining} used an anonymous survey to explore both the self-reported characteristics and motivations amongst users of booter and stresser services. They contacted some 59,009 account holders and gathered partial responses from 821 individuals, with at least 250 responses for each question. Although this provided sufficient data for analysis and represents a reasonably large survey sample, it was still a very small proportion of the number of people initially contacted who chose to provide data. Overall, self-reported characteristics included (a) that respondents were `primarily' in their late teens or 20s, (b) that `the majority' reported living in the U.S. or U.K., and (c) that many were educated beyond high school level. 88\% of respondents also reported being male, and 63\% reported being white. Finally, the survey also asked respondents to rate their skill level on a scale from 1 to 10, with 1 being the lowest and 10 being the highest. The most frequent response was 10, followed by 1, which suggests either a substantial divergence in actual skill level or in the degree of confidence that individuals have (e.g., displaying high levels of over- and under- confidence).  

Although published in 2003, \citet{woo2003hacker} was not included in previous reviews, and covers a significant population of cybercriminals. The author conducted an anonymous survey with some 729 self-reported hackers examining a range of socio-demographic factors. 88\% of respondents were found to be male, with only 4.3\% being female, and the remainder not identifying their sex. Hackers were young, with over half being younger than 19, roughly 83\% being younger than 25, and 98\% being younger than 35. About 40\% of hackers reported no religion, followed by Christianity (24.7\%), Buddhism (19.3\%). Reported ethnicity was shown to be heavily Asian (60.4\%), followed by Caucasian (18.8\%). This is distinct from other literature suggesting a majority of cybercriminals are white, likely due to the survey being distributed in Korean as well as English. First languages were highlighted as Korean (48.6\%), English (29.2\%), Afghan (8.4\%), Spanish (1.5\%) and Japanese (1.4\%).
Respondents conformed to perceptions of hackers as mostly educated. The greatest number had, at most, completed 4 years of college [university] (26.6\%), followed by high school (25.2\%), elementary school (15.2\%), 2 years of college (13.9\%), middle school (11.8\%), master's degree (4.7\%) and Ph.D. (2.6\%). 
When considering the types of hacking that individuals had engaged in, respondents reported having broken into personal sites (70\%), university web sites (56.5\%), small business web sites (47.2\%), big business websites (42.2\%), foreign government sites (38\%), porn sites (35.9\%), foreign ethnicity websites (31.8\%), military websites (27.8\%), transnational corporation sites (27.7\%), secret agency sites (26.5\%), foreign religion websites (22.6\%) and bank computer systems (21.9\%). However, since these claims can not be verified, the extent to which this information is accurate is unclear, and may represent a higher estimation due to potential over-confidence in ability or a desire to communicate a high status or  expertise.

\citet{donner2016gender} conducted an online survey of 522 students enrolled on social science or general education courses in the US, examining self-reported engagement in three categories of cyber crime (hacking, digital piracy and cyber-harassment) and the potential relationship with individual characteristics, such as gender, degree of self-control, and degree of immersion in the cyber environment. Regression analyses conducted on the data suggested that men were more likely to engage in hacking activities than women, however this relationship was dependent on the degree to which individuals used the Internet, with the gender difference disappearing among high Internet users. Conversely, being older and spending less time online was associated with lower engagement in hacking activities. This suggests that findings related to greater involvement in deviant online activities, such as hacking, by males may in fact be due to a traditionally greater propensity of Internet and computer use by such individuals.

\citet{pyrooz2013gangs} interviewed 585 young adult respondents across different cities in the US about their use of the Internet, their involvement in gangs, and their criminal and deviant activities online. Overall, 45\% of the sample were found to have engaged in some form of online offending in the last 6 months (e.g., selling drugs, digital piracy, selling stolen property) and both gang membership and being younger was found to be associated with increased online offending. Greater technological competence, spending a greater amount of time on the Internet, and greater use of social networking, were also all found to be associated with greater online offending.

\citet{liao2017computer} studied data on offenders from a national reporting system, examining differences in arrest likelihood according to the form of cybercrime engaged in by some 595 offenders, reporting also on offender characteristics. Their findings raise questions for studies based on arrest data or police reports. They found that personalised fraud (confidence tricks, impersonation) has a significantly lower likelihood of arrest compared to impersonal credit card fraud. Greater age was associated with greater probability of arrest, with law enforcement treating the many offenders younger than 20 with more leniency. Females were more likely to be arrested than males.

\citet{steinmetz2015becoming} took an ethnographic approach to understanding hackers, attending public in-person meetings for a hacker group. Ethnographic observations on demographics and social background were supplemented with 14 semi-structured interviews. Considerable ethnographic detail is presented about the developmental factors, parental influence, exposure to technology and exposure to hacking. Support was found for a gender and race disparity, but not for a youth bias. Most participants considered themselves middle-class, and were employed in the technology sector. A parental involvement in a technological occupation was also common, with Steinmetz positing that early exposure to technology is followed by a drift toward hacking in adolescence, which may then be confirmed through self-identification.

Examining cases of known offenders who were prosecuted by law enforcement, \citet{payne2018using} undertook a content analysis of 119 cybercrime cases prosecuted by the US Dept of Justice between 2013 and June 2017 and found that the most common cases related to hacking and online fraud, with 91.5\% of offenders being male. Interestingly, the average age for male offenders (mean age 35.6 years) was found to be younger than for female offenders (mean age of 51.5 years). More than half of the cases were also found to involve more than one offender, suggesting that either such online activities have an important social or group-based component, or that offenders who work with others are merely more likely to be identified and prosecuted. The research also found that 30\% of offenders were from countries outside the US and that, interestingly, these offenders tended to be younger than US offenders.

In their survey of 488 university students in the US, \citet{donner2014low} also considered the role of degree of self-control, age and degree of Internet use in influencing different forms of deviant online activities (including hacking, distributing malware, and  using the internet for a drug transaction). Correlation and regression analyses conducted on this data found that low self-control was associated with the majority of online deviance (specifically, 7 out of the 10 forms of online deviance). Greater Internet use was also associated with greater online deviance overall, supporting both self-control theory (i.e., that greater criminal deviance is associated with a decreased ability to control behaviour in the face of temptation) and routine activities theory (i.e., that greater presence in an environment where opportunities for criminal activities are available is associated with greater criminal activity). However, the latter still requires the offender to be sufficiently motivated and for there to be sufficient opportunities for engagement in criminal activities, which are not necessarily an obvious result of greater Internet use per se. Similarly, in their survey of 2751 Korean youths who were part of a longitudinal youth survey panel and had an average age of 14 years, \citet{moon2010general} found that low self control was associated with a greater likelihood of illegally downloading software and a higher likelihood of illegally using others' personal identity online. Greater time spent using computers, joining a cyber club, being male, and having greater academic competence were also associated with illegally downloading software.

\citet{seigfried2015assessing} undertook an online survey with 296 respondents who were undergraduate students from a variety of disciplines at a Southern US university to explore the relationship between various characteristics associated with Asperger's Syndrome (which was assessed using the self-report Autism Quotient scale) and a range of deviant computer activities, including hacking and virus writing. Surprisingly, hacking was found to be negatively associated with the AQ subscale attention to detail, suggesting that lower attention to detail was related to increased self-reported engagement in hacking activities. Identity theft was also found to be positively associated with overall score on the AQ, as well as poor social skills, poor communication and poor imagination, with virus writing similarly positively associated with overall AQ score, poor social skills, poor communication and poor imagination. Overall, greater levels of computer deviance in general were associated with higher total AQ scores, as well as particular sub-scale scores related to social skills, communication and imagination.

Finally, \citet{chan2014exploratory} used an online auction offender database containing case reports collected between 2012 and 2013 in Hong Kong (61 case reports in total) to examine various socio-demographic characteristics of offenders. Variables from these case reports (e.g., age, education level etc) were divided into offender behavior and personal characteristic types. Three clusters of offender were then identified, with age, education level and intrinsic motivation (both objective and why [an offender was] not afraid of being caught) identified as being primarily responsible for differentiating between these different offender types. Interestingly, intrinsic motivation and age was suggested to be the most evident aspects for experienced-active offenders.

\newpage
\section{Motivations of Cybercriminals}
\label{sec:motivations}

\begin{mdframed}[style=SummaryBox]
\begin{itemize}
	\item A number of existing criminological theories, such as Neutralisation Theory and Self-Control Theory were applied to the cybercrime context. 
	\item Application of other approaches, such as the Theory of Planned Behaviour, is increasing, but with gaps for exploitation.
	\item A number of studies examine the overlap between prior victimisation and offending, which is considerable. 
	\item Distinctions between different kinds of cybercriminal could explain differing motivations.  
	\item A summary of motivations supported by current and previous studies is presented in Table~\ref{tab:motivations}.
\end{itemize}
\end{mdframed}

The various factors that may motivate individuals to engage in different types of cyber-dependent crimes has been the focus of increasing research in recent years. Such motivations may range from financial reward and organised criminal set-ups that mirror legitimate business sectors, to individual vengeance, political statements, and the pursuit of excitement, status and intellectual challenge. This section reviews research investigating these motivations, with Table~\ref{tab:motivations} providing a summary of findings in this area to date.

\citet{xu2013computer} provide a review of literature on cybercrime motivations prior to 2013, including a number of theoretical lenses such as social learning theory, rational choice theory and neutralisation theory. They also present a case study of six Chinese hackers, discussing elements from their histories such as early interest in computers, innocent motives for hacking, being under-challenged by education, endemic poor computer security, tolerance by schools and peer support. They found general support for routine activity theory, social learning theory and situational action theory, and provide an integrative framework for combining the three theories in a developmental model of hacker evolution.

Although conducted in 2003, \citet{voiskounsky2003flow} undertook an interesting study exploring the experience of flow within Russian-language hackers. Overall, hacker motivations were considered, at that point, to be largely intrinsic and cognitive (i.e., reflecting internal mental processes) and the researchers focused on investigating the presence of `flow' as a causative factor, expecting to find that it occurs more often in more qualified and competent hackers. The psychological process of flow relates to being fully immersed in an activity to the extent that it completely absorbs the individual and is associated with enjoyment and energised focus. This immersion is known as experiencing a \textit{flow state.} In their survey of 457 self-reported Russian-language hackers, who were drawn from online venues of different competence levels, the researchers showed that greater reported flow was associated with the less-competent population. Analysis of follow-up interviews also suggested that such flow can be interrupted in a `zigzag' fashion, with crises related to lack of skill motivating hackers' further development. 

In their 2003 review of previous work categorising and understanding the motives of ideological or hobby-driven hackers, \citet{woo2003hacker} approached the topic theoretically by considering factors related to self-esteem (particularly narcissism), extrinsic (i.e., external) vs. intrinsic (i.e., internal) motivations, flow, and threats to the hacker's worldview. A survey was distributed at an online hacking event in English and Korean and 729 valid responses were collected and analysed. These showed an association between hackers' narcissism and aggressiveness. Higher extrinsic motivation in hackers was associated with aggressive temperament, and specific categories of hacking target. Hacking was also positively associated with the experience of flow, with further analysis showing that this was overall positively associated with intrinsic-motivation, and negatively related to extrinsic motivations. Finally, higher nationalism and religious pride scores were associated with angry temperament.

\citet{broadhurst2014analysis} covered the diverse motivations of groups involved in known cases of cybercrime of various kinds. Individual offenders ``appeared less preoccupied with financial gain than with libertarian ideology, technological challenge, celebrity obsession, and revenge against a former employer''. Similarly, organisations were not always focused on financial gain, and ``reflected a variety of goals, including defiance of authority, freedom of information, sexual gratification of members, and technological challenge''. However, the profit motive was more apparent in the organization cases than with individual offenders. Relatively complex protest activity and `annoyance crime' such as that involving denial of service attacks, seemed to be best served by a `swarm' organization structure, with protest activity being of a more ad hoc, short-term nature.

More recently, \citet{hyslip2018defining} asked users of booter and stresser services about their motivations, with 230 responses. The majority reported innocuous use of the services for system testing (60\%), research (49\%) or to test their own abilities (47\%). Others reported using the services to affect an online game or game opponent (41\%), and for hacktivism purposes (26\%). Smaller categories reported being hired to use the services (19\%) or using them to affect a business competitor (11\%). Many users (54\%) reported running their own similar services, and 76\% reported having been targeted by such a service themselves in the past.  An earlier and much smaller study of 11 operators of booter and stresser services by \citet{hutchings2016exploring} provides deeper qualitative insight into their motivations. Broad support was found for rational choice theory (the financial gains of users) and neutralisation theory (their focus on legitimate usage of these systems).

Prior victimisation is often encountered as a motivation. \citet{kerstens2016victim} studied the overlap between victims and perpetrators of three categories of financial cybercrimes (auction fraud, virtual theft and identity fraud) in a cross-sectional study of 6,299 Dutch youths aged 10 to 18. Victimhood and perpetration were assessed through self-report items in an online questionnaire, along with measures of the respondents' online behaviour, parental mediation and self-control. The overlap between victims and perpetrators was significant, with victims more than six times as likely as the general population to report perpetrating a cybercrime, and perpetrators more than five times as likely to report victimhood. Motives reported for perpetration included financial gain, entertainment and bullying as well as retaliation. A multinomial regression analysis found significant effects for both victimhood and perpetration for being male, having a high frequency of internet usage, a higher level of online disinhibition, a higher level of online disclosure, and a lower level of self-control.

Building on this result in more detail, \citet{kranenbarg2017offending} compares victimisation, offending and victim-offending between 240 cybercrime suspects and 219 traditional crime suspects. They find a considerable victim-offender overlap in cybercrime, but this is qualified by the technical sophistication of attacks. For instance, hacking by guessing a password was more common amongst victim-offenders, while other forms were more often committed by pure offenders. Pure offenders were more likely to spend time on forums and possess greater IT skills. Low self-control was an important predictor for being a cybercrime victim-offender (as opposed to victim or offender only). IT skills and spending more time on online shopping also increased the likelihood of victimization-offending, as well as living with parents (though this effect was reversed amongst pure offenders). 

\citet{marcum2014hacking} studied self-reported offending in a survey of 1,617 high-school students, measuring low self-control and association with deviant peers. Logistic regression showed that low self control and deviant peers were associated with logging into Facebook accounts and accessing websites without permission. Deviant peer association was associated with sending email as another person without permission, but not low-self control. A number of demographic variables were also found to be predictive of particular categories of deviant behaviour.

\citet{ho2012influences} conducted a survey of 1644 12-17 year olds in Taiwan who were recruited via schools and correctional institutions and suggested that gratification and previous victimisation experiences were amongst the most influential factors in self-reported engagement in deviant computer activities. In their MSc thesis, \citet{mcbrayer2014exploiting} used snowball sampling to recruit 120 survey respondents to explore motivational factors for different types of cyber crime behaviours. Overall, it was found that males were more likely to be script kiddies, cyberpunks, and old guard hackers compared to females and that these computer deviant behaviors overlapped in both motivational factors and the behaviors themselves. Script kiddie, password
cracker, and old guard hacker behaviors were found to be  motivated primarily by addiction processes, whereas cyberpunk behavior was found to be motivated by financial, peer recognition, and revenge motivations, and internal computer deviant behavior was related to financial and peer recognition motivations.

In their report exploring cybercriminal overlap with traditional organised crime \citet{odinot2016cybercrime} examined 11 criminal cases and 107 suspects, 39 of which had roles connected to ICT use. The investigation found four broad categories of criminality: fraud relating to banks and payment systems; trade in illict goods (including online drugs sales, botnet sales); extortion (ransomware, DDoS extortion) and data theft. Case files revealed that criminal motives included making money, paying off debts, revenge, coercion by others, and `as a hobby'.

In 2018, \citet{kranenbarg2018cyber} examined longitudinal registration data on adults suspected of cybercrimes (870) or traditional crimes (1,144,740) in the Netherlands between 2000 and 2012. They examined a number of lifestyle patterns suggested to reduce the likelihood of committing crime, including whether the suspect was living with a romantic partner or child, whether they were employed, and whether they were enrolled in education. They found that living with a partner and child decreases the likelihood of committing cybercrime by 46\%, compared to a decrease of 19\% for traditional crime. Having a job also reduced the likelihood of offending by 10\% (compared to 7\% for traditional crime), but if the job was IT-related, likelihood was actually \textit{increased} by 14\% (compared to a decrease of 11\% for traditional crime). This is likely due to increased computer knowledge and increased computer and Internet usage presenting a risk factor, as highlighted by research discussed in Section 2. Education level was not found to produce any significant effects, but the direction of effects tentatively suggested that greater education may increase the likelihood of engaging in cybercrime offences.

Finally, \citet{holt2017exploring} conducted an online survey of 779 students in a midwestern university in the US exploring the different factors that may influence the likelihood that an individual will engage in the defacement of websites. Regression analyses showed that computer ownership and not feeling that there are clear rules about online behavior were associated with being more willing to engage in website defacements. In addition, respondents who felt that law enforcement didn't respond to cybercrime and those who were also more willing to engage in offline protest behaviors were more willing to deface websites. 

\begin{center}
\footnotesize
\begin{longtable}{l l}
\caption{Table of supported motivations reproduced from \citet{mcbrayer2014exploiting} for pre-2013 evidence, and updated with reference to the studies discussed above.}\\
\hline
Motivation & Supporting Studies \\
\hline
\endfirsthead
\hline \multicolumn{2}{|r|}{{Continued on next page}} \\ \hline
\endfoot
\hline
\endlastfoot
\multicolumn{2}{c}{{\tablename\ \thetable{} -- continued from previous page}} \\
\hline
\endhead

\multirow{12}{*}{Addiction/Low Self Control}	& \citet{taylor1999hackers} \\
				& \citet{woo2003hacker} \\
				& \citet{beveren2001conceptual} \\
				& \citet{voiskounsky2003flow} \\
				& \citet{turgeman2005hackers} \\
				& \citet{smyslova2009usability} \\
				& \citet{marcum2014hacking} \\
				& \citet{mcbrayer2014exploiting} \\
				& \citet{donner2014low} \\
				& \citet{moon2010general} \\
				& \citet{kerstens2016victim} \\
				& \citet{kranenbarg2017offending} \\
\hline

\multirow{7}{*}{Curiosity} 	& \citet{beveren2001conceptual} \\
				& \citet{voiskounsky2003flow} \\
				& \citet{woo2003hacker} \\
				& \citet{turgeman2005hackers} \\
				& \citet{smyslova2009usability} \\
				& \citet{nikitina2012hackers} \\
				& \citet{xu2013computer} \\

\hline
\multirow{11}{*}{Excitement/Entertainment} & \citet{beveren2001conceptual} \\
					& \citet{gordon2003convergence} \\
					& \citet{voiskounsky2003flow} \\
					& \citet{turgeman2005hackers} \\
					& \citet{smyslova2009usability} \\
					& \citet{ho2012influences} \\
					& \citet{xu2013computer} \\
					& \citet{chan2014exploratory} \\
					& \citet{kerstens2016victim} \\
					& \citet{odinot2016cybercrime} \\
					& \citet{broadhurst2014analysis} \\
\hline
\pagebreak
\multirow{12}{*}{Money} & \citet{taylor1999hackers} \\
			& \citet{gordon2003convergence} \\
			& \citet{woo2003hacker} \\
			& \citet{kilger2004profiling} \\
			& \citet{turgeman2005hackers} \\
			& \citet{holt2007subcultural} \\
			& \citet{chan2014exploratory} \\
			& \citet{mcbrayer2014exploiting} \\
			& \citet{broadhurst2014analysis} \\
			& \citet{kerstens2016victim} \\
			& \citet{odinot2016cybercrime} \\
			& \citet{hyslip2018defining} \\

\hline
\multirow{9}{*}{Power/Status/Ego} 	& \citet{taylor1999hackers} \\
					& \citet{woo2003hacker} \\
					& \citet{holt2007subcultural} \\
					& \citet{nikitina2012hackers} \\
					& \citet{xu2013computer} \\
					& \citet{broadhurst2014analysis} \\
					& \citet{mcbrayer2014exploiting} \\
					& \citet{marcum2014hacking} \\
					& \citet{kerstens2016victim} \\

\hline
\multirow{7}{*}{Ideology}	& \citet{woo2003hacker} \\	
				& \citet{kilger2004profiling} \\
				& \citet{schell2004cybercrime} \\
				& \citet{jordan2004hacktivism} \\
				& \citet{holt2009attack} \\
				& \citet{broadhurst2014analysis} \\
				& \cite{hyslip2018defining} \\

\hline
\multirow{17}{*}{Revenge}	& \citet{taylor1999hackers} \\
				& \citet{loper2001criminology} \\
				& \citet{voiskounsky2003flow} \\
				& \citet{best2003hacker} \\
				& \citet{kilger2004profiling} \\
				& \citet{turgeman2005hackers} \\
				& \citet{williams2006virtually} \\
				& \citet{holt2007subcultural} \\
				& \citet{mcquade2008encyclopedia} \\
				& \citet{ho2012influences} \\
				& \citet{xu2013computer} \\
				& \citet{mcbrayer2014exploiting} \\
				& \citet{broadhurst2014analysis} \\
				& \citet{kerstens2016victim} \\
				& \citet{odinot2016cybercrime} \\
				& \citet{kranenbarg2018cyber} \\
				& \citet{hyslip2018defining} \\
\hline
\label{tab:motivations}
\end{longtable}
\end{center}

\section{Quality Assessment}
\label{sec:quality}


Research included in this review was subjected to a quality assessment to enable the potential contribution, generalisability and likely replicability of the research to be evaluated. In this assessment, we have focused on considering the participant sample size and sample type, the publishing outlet (e.g., the impact factor of academic journals), and whether sufficient information is provided within the research methodology of the article to enable the method to be effectively reproduced by other researchers if they had access to a comparable sample population. As cybercriminality appears to be a rapidly-evolving subject, and older work may quickly lose relevance, we also note the year of publication for studies.

The overall sample size for a study ($N$) and the sample of cybercriminals ($C_n$) are recorded separately. Many studies surveyed larger populations looking for evidence (usually self-reported) of cybercriminality. Understanding the strength of the evidence about cybercriminals requires both figures to be considered. The definition of cybercriminality used was that adopted by each study, so samples may not be directly comparable. The type of sample used was coded based on the method presented in each study. Impact factors were retrieved from journal homepages where available. Replicability was coded as one of: \begin{itemize}
	\item[\bf{n}] Significant missing information about the method, or the method is qualitative by nature; 
	\item[\bf{m}] Most detail about the study approach is available, but specifics of questions/measures used are missing or incomplete;
	\item[\bf{y}] Sufficient detail for full replication, given access to a comparable population.
\end{itemize}

\begin{table}[ht]
\footnotesize
\tabcolsep=0.06cm
\centering
\caption{Quality analysis elements for included studies. $N$ = Total sample covered; $C_n$ = Total cybercriminals covered; IF = journal impact factor; $\varnothing$ = figures not reported (may be subcategories or other measures); * = identified by police report.}
\begin{tabular}{lrrlrcr}
  \hline
Paper & $N$ & $C_{n}$ & Type & IF & Replicable & Year \\ 
  \hline
\citet{steinmetz2015becoming} &  14 &  14 & cybercriminal cases &  & n & 2015 \\ 
  \citet{xu2013computer} &   6 &   6 & cybercriminal cases  & 3.06 & n & 2013 \\ 
  \citet{broadhurst2014analysis} &   6 &   6 & cybercriminal cases &  & n & 2014 \\ 
  \citet{hyslip2018defining} & 821 & 821 & cybercriminal survey &  & m & 2018 \\ 
  \citet{woo2003hacker} & 729 & 729 & cybercriminal survey &  & y & 2003 \\ 
  \citet{voiskounsky2003flow} & 457 & 457 & cybercriminal survey & 2.69 & m & 2003 \\ 
  \citet{hutchings2016exploring} &  11 &  11 & cybercriminal survey & 1.05 & m & 2016 \\ 
  \citet{mcbrayer2014exploiting} & 120 & 112 & cybercriminal survey &  & y & 2014 \\ 
  \citet{kranenbarg2017offending} & 535 & 268 & cybercriminal survey* & 1.05 & y & 2017 \\ 
  \citet{liao2017computer} & 595 & 595 & police reports & 3.23 & y & 2017 \\ 
  \citet{payne2018using} & 119 & 119 & police reports & 0.95 & m & 2018 \\ 
  \citet{chan2014exploratory} &  61 &  61 & police reports & 5.06 & m & 2014 \\ 
  \citet{kranenbarg2018cyber} & 1,145,610 & 870 & police reports &  & y & 2018 \\ 
  \citet{odinot2016cybercrime} & 107 & 39 & police reports &  & m & 2016 \\ 
  \citet{donner2016gender} & 522 & $\varnothing$ & student survey &  & y & 2016 \\ 
  \citet{donner2014low} & 488 & $\varnothing$ & student survey & 3.54 & y & 2014 \\ 
  \citet{seigfried2015assessing} & 296 & 179 & student survey & 1.38 & y & 2015 \\ 
  \citet{marcum2014hacking} & 1,617 & $\varnothing$ & student survey & 1.05 & y & 2014 \\ 
  \citet{ho2012influences} & 1,592 & $\varnothing$ & student survey &  & n & 2012 \\ 
  \citet{holt2017exploring} & 779 & $\varnothing$ & student survey & 1.05 & y & 2017 \\ 
  \citet{pyrooz2013gangs} & 585 & 209 & youth survey & 2.22 & y & 2013 \\ 
  \citet{moon2010general} & 2,751 & 473 & youth survey & 3.14 & y & 2010 \\ 
  \citet{kerstens2016victim} & 6,299 & 862 & youth survey & 1.05 & y & 2016 \\ 
   \hline
\end{tabular}
\label{tab:quality}
\end{table}

Table~\ref{tab:quality} reports on the chosen quality analysis measures for the papers included in this review, grouped by sample type. Most studies were surveys of student or youth populations, or online surveys of self-identified cybercriminals. \citet{kranenbarg2017offending} is notable for surveying cybercriminals identified by police reporting information (suspects in a case). Other uses of police report information tended to work from case file details and national registers.

Overall sample sizes ($N$) were varied, ranging from discussion of six cases~\citep{xu2013computer} to analysis of over a million offenders of any type in a given timeframe~\citep{kranenbarg2018cyber}. The median overall sample size was 522. Samples of cybercriminals ($C_n$) were usually smaller and varied less widely (median=194; mean=324; sd=324). Several studies reported on particular subcategories or measures of cybercriminality in their sample in a manner such that the overlap between subcategories could not be established in order to extract a single figure for the size of cybercriminal population studied. These studies (all student surveys) are marked as $\varnothing$ in Table~\ref{tab:quality}.

Replicability of studies was reasonable, with only four studies marked as \textbf{n}, the lowest level, and three of these being explicitly qualitative in their methodology. Over half (13) of the studies were graded as \textbf{y}, highly replicable given access to a comparable population. Student and youth surveys tended to be more often highly replicable than studies of police reports or online surveys of cybercriminals, perhaps due to differing backgrounds of authors. 

The average journal impact factor was 2.181. However, journal impact factors were not available for many (9) studies. In addition, over one-third of the studies for which IF was available originated from the same journal -- \emph{Deviant Behaviour} -- which may be given undue weight as a result.

\section{Limitations and Current Research Gaps}
\label{sec:limitations}

\begin{mdframed}[style=SummaryBox]
\begin{itemize}
	\item Research is limited by poor access to cybercriminal communities. Self-report methodologies are highly common.
	\item Studies in students or youth populations make presumptions about cybercriminality, and may not reflect technically sophisticated, financially-motived or organised cybercrime.
	\item Varying and sometimes loose definitions of cybercriminality, cyber-deviance or ``hacking'' compound comparison problems.
	\item Priorities for improvement should include identifying data sources for diverse samples of technically-sophisticated cybercriminals, avoiding self-report structures.
\end{itemize}
\end{mdframed}


Although the current research base investigating the primary motivations and characteristics of cyber criminals continues to increase, work in this area remains limited. Existing research is also constrained by methodological limitations, predominantly due to difficulties in accessing offender populations and a reliance on self-report measures and self-selected samples. However, the growing use of techniques that enable the scraping of forums and the analysis of large volumes of data (e.g., data mining, machine learning) provide a mechanism to continue to improve and develop current understanding of cybercriminal behaviour. This section discusses the limitations of previous research in this area, including the primary research methods used (e.g., self-report surveys, case data, interviews) and the potential issues associated with these. This section also highlights current gaps within the existing research base, which may limit the extent to which robust conclusions can be drawn, and emerging methodological approaches that aim to address some of these issues. Table~\ref{tab:quality} above highlights the key methodological details of identified research and associated limitations. 

The majority of research in this area has used self-report methods (including online surveys and less commonly, in-person semi-structured interview approaches). This has provided a range of data types, including qualitative themes identified via responses to open-ended questions, and quantitative data that commonly uses responses to questionnaire measures addressing demographics, psychological characteristics (such as self-control) and reported engagement in a number of online activities. This data is then typically analysed using regression approaches to determine which factors predict cyber criminal activities. Online surveys often utilise student samples, or existing research panels that are available to access, particularly when attempting to investigate participants from younger demographic samples, which does lead to limitations with regards to the generalisability of findings across wider population groups. Work focused on accessing other groups, including law enforcement or other forms of debriefs and data collected via volunteers on forums, provides other mechanisms to increase the generalisability of research findings, but bring further limitations (e.g., individuals who have been caught/ are willing to engage with debriefs, self-selected samples within volunteers). 
While online surveys often capture international cybercriminals, and there are studies focused on Russian, Chinese and Korean cybercriminals, the majority of studies are focused on cybercriminality in the US, UK and Netherlands, making international comparisons difficult. 

Other methods that are increasingly used include analysing data that has been scraped from online forums, such as \citep{holt2013examining,holt2013exploring,motoyama2011analysis,decaryhetu2013reputation}, which to date commonly focus on market dynamics and the impact of law enforcement interventions rather than offender motivations or characteristics. There is also the examination of available data on prosecutions, arrests, and the characteristics associated with this by some researchers. This can also include a case-based theoretical consideration of potential factors that are identified across different case reports and how these relate to existing criminological and psychological theory. 

Much of the prior work on cybercriminal characteristics and motivations has suffered somewhat from a lack of distinction between different categories and subcategories of cybercriminal offending, with some authors noting this explicitly.
\citet{rogers2006two} provides a taxonomy of ``hackers'' which attempts to address the confusion between different categories of offending, positing eight or nine distinct classifications with their own distinctive features: 
 \emph{Novices}, who rely on pre-written tools for thrill-seeking or ego stroking purposes; 
 \emph{Cyber-punks}, who have limited programming ability which they employ for attention or financial gain; \emph{Internals}, who are disgruntled (and possibly ex-) employees seeking revenge;
 \emph{Petty Thieves}, who are traditional criminals following their targets into the digital realm;
 \emph{Old Guard}, who have deep technical skill and hack as an intellectual exercise;
 \emph{Professional Criminals}, who are professional financially-motivated criminals;
\emph{Information Warriors}, who are state-sponsored combatants in the digital realm;
\emph{Political Activists}, who are ideologically motivated for notoriety; and 
\emph{Virus Writers} -- the background of which is poorly distinguished. 
This taxonomy gives a frame of reference for understanding the targeted population of some empirical studies of cybercriminals. However, many studies reviewed use far looser definitions of cybercriminality or `cyber deviance' than any of the above categories. Particularly in student and youth surveys, but also in analyses of police case reports, there is a lack of distinction between low-level juvenile cybercriminality involved in, e.g., guessing a person's Facebook password and accessing their account without permission, versus deploying skilled technical attacks or taking part in organised cybercrime. This broadened definition may be motivated by the difficulties in accessing cybercriminals of the latter type.

Several researchers have deployed anonymous online surveys to investigate the background and motivations of cybercriminals. These have plausibly better access to this latter category of advanced cybercriminality, although with the penalty of far less verifiable data. Some demographic details from these surveys have proven relatively consistent: for example, that the population is predominantly young~\cite{hyslip2018defining,voiskounsky2003flow,woo2003hacker,seigfried2015assessing} and male~\cite{hyslip2018defining,woo2003hacker,payne2018using}. While the heavy gender bias is also confirmed in results from surveys carried out at hacker conferences~\cite{bachmann2010risk,young2007hacking}, and other sources~\cite{hutchings2016exploring,payne2018using} the evidence on age may be skewed by targeting \emph{Novice} or \emph{Cyber-punk} venues (\citet{hyslip2018defining} target users of stresser services, and \citet{woo2003hacker} surveyed participants in an esteem-driven hacking contest). Some authors have challenged the perception of cybercriminals as mostly young~\cite{steinmetz2015becoming}, suggesting a middle-aged population, which may reflect a different category of offender. This is a significant limitation for generalisation from the (dominant) youth and student surveys, which necessarily exclude older cybercriminals. Other demographics have proven more sensitive, for example, the majority ethnicity of respondents differing apparently according to the languages in which surveys were available (cf. \cite{woo2003hacker,hyslip2018defining}).

While regressions on survey results have pointed towards potential predictors of hacking such as narcissism, nationalism and propensity toward `flow' states~\cite{woo2003hacker}, the accuracy of survey self-reports about skill or criminal accomplishments is uncertain and somewhat suspect, with, e.g., 26\% of \citet{woo2003hacker}'s respondents reporting having broken into secret agency sites, and the most common self-reported skill rating on \citet{hyslip2018defining}'s 10-point scale being 10. \citet{voiskounsky2003flow} attempt to control for this by delivering their survey to two Russian forums of reportedly different quality levels, and using simple tests of the depth of IT knowledge in respondents, with some apparent success, finding that `flow' is inversely predictive of skill when measured in this manner.  

Other approaches for understanding cybercriminals have worked from case details drawn from police investigations, often producing more qualitative insight into the operation of cybercrime. \citet{leukfeldt2017online}, for example, describe the function of online marketplaces as exchanges and enablers of crime with reference to some 40 internationally-sourced cases. \citet{odinot2016cybercrime} use 11 police files to explore the link between cybercrime and traditional organised crime, finding interesting deviations in offender age profile which may be influenced by the balance of traditional organised crime, and motives including financial, revenge and hobby-driven cases. Case files are illustrative, but comparatively few, and may not capture the full nature of ongoing cybercrime. Even within cases, the range of suspects included may be more strongly influenced by the focus of a police inquiry than the nature of the offender community~\cite{odinot2016cybercrime}.

Although the above aspects present a range of limitations that should be considered when evaluating the relative robustness of findings in this area, future work should also consider ways to address these limitations and thus reduce current research gaps. Although it is appreciated that such work represents a substantial challenge, current efforts should focus on:
\begin{enumerate}
\item Identifying data sources that provide access to a more diverse sample, such as individuals from various socio-demographic backgrounds and geographic locations, as well as encompassing a range of types of activity that have been / are currently engaged in and a range of knowledge levels and expertise. 
\item Identifying data sources that do not rely on self-report responses or self-selection of participants, using objective sources of data where possible. Such an approach may require a more systematic collection of different types of data by certain bodies, and a greater ability to share such data with researchers where possible.
\item Developing methods of analysis that are less resource-intensive than traditional qualitative approaches and are able to reliably analyse large quantities of both linguistic, and other, data types. 
\end{enumerate}

The AMoC project aims to provide a means to address some of these aspects, developing reliable, evidence-based methods to automatically analyse large volumes of textual data gathered from a range of online sources.
More specifically, previous work by \cite{rao2010classifying,peersman2011predicting,nguyen2013old,al2012homophily,rashid2013analyzing,rangel2015overview} has demonstrated that automatic user profiling in online social media based on their text messages can be highly accurate, even when only limited data per user is available and in the presence of adversaries (i.e., users who attempt to hide their true identity or who create a false digital persona) \cite{rashid2013analyzing,peersman2018detecting}. Language can also be identifying of location, with language identification and native language identification techniques helping to narrow down the origin of users \cite{malmasi2017multilingual,malmasi2017report,malmasi2015large}.
Within the AMoC project, an evaluation of the efficiency of existing text mining approaches will be evaluated to assess their forensic readiness when applied in the context of forensic investigations pertaining to cybercrime. Additionally, novel, advanced techniques to automatically identify cyber offenders and their characteristics (demographic, psychological and situational) across different online environments based on their linguistic fingerprint will be developed.

\newpage
\section{Conclusions}
\label{sec:conclusions}

This review aimed to evaluate current research regarding the motivations and characteristics of individuals engaging in cyber-dependent crimes, focusing on research conducted within the last 7 years (i.e., 2012 onward). Overall, a range of literature has been identified that provides a degree of support for a number of primary characteristics and motivations, which have been examined across a range of studies and have been found to be sufficiently supported in the prior literature to warrant inclusion in an initial framework. However, it should be noted that these findings may reflect potential methodological biases in accessing various population samples in prior studies. It also does not suggest that individuals outside of these characteristics will not be engaged in cyber criminal activities. These characteristics include:\\

\begin{mdframed}[style=CharBox]
\begin{itemize}
    \item \textbf{Age:} A greater likelihood of being in relatively younger age groups, particularly individuals in their late teens and twenties;
    \item \textbf{Gender:} A greater likelihood of being male;
    \item \textbf{Education level}: A greater likelihood of completing a higher level of education (e.g., high school and beyond) and greater academic competence;
    \item \textbf{Extent of internet use/ computer use:} Spending a greater amount of time on the internet and engaging in a greater degree of computer use, as well as greater technological competence associated with this;
    \item \textbf{Employment status:} Having a greater likelihood of being employed in computing and technology occupations or of being unemployed;
    \item \textbf{Trait self-control:} Having a greater likelihood of being low in trait self-control.
\end{itemize}
\end{mdframed}
\newpage
Motivations that have been highlighted within the current research literature include:
\begin{mdframed}[style=MotivBox]
\begin{itemize}
\item financial reward; 
\item a desire for status, power and recognition from others;
\item addiction processes and other forms of hedonic gratification; 
\item a desire for vengeance against particular individuals, groups or society;
\item curiosity related to accessing particular information or information sources, or the technical protections related to them;
\item immersion in an activity, experiencing psychological flow during the process of engaging in the activity, and developing skills and expertise in relation to particular activities;
\item previous victimisation in relation to cyber crime, exposure to cyber criminal activities and the people that engage in them;
\item risk perceptions related to the perceived likelihood of being caught and punished for engaging in cyber criminal activities;
\item the perceived social  and moral acceptability of engaging in cyber criminal activities (e.g., the perception that there are no clear rules governing online environments or no real victims of cyber crime). 
\end{itemize}
\end{mdframed}

However, the majority of this research was found to focus on: (\textit{i}) the use of self-report methods, exploring what have historically been more common deviant online behaviours, such as digital piracy and, to a degree, hacking activities; and (\textit{ii}) applying established criminological theories, such as Neutralisation Theory and Self-Control Theory, to the cyber crime context. Hence, the research surveyed in this review still shows a number limitations. Self-report methods remain restricted by self-selection bias (i.e., only people relatively easy to access and interested in taking part have completed surveys), and a difficulty in validating people's responses (i.e., have they really engaged in the behaviours they claim, for the reasons that they claim? Are they potentially under- or over-reporting certain aspects?). Additionally, although the application of other psychological / behaviour change theories, such as the Theory of Planned Behaviour, is increasing, there remains an opportunity to undertake further work considering these approaches in relation to cyber criminal behaviour. As a result, these factors will be used to provide an evidence-based foundation for the future work of the AMoC project, in order to explore whether these concepts are also supported when investigated in relation to their presence in objective, forum-based data taken from a range of forum sites and which historically include individuals from a range of geographic locations and backgrounds. 

This data is being gathered by the project team and will be analysed using a combination of quantitative automated approaches and more nuanced qualitative approaches using human coders. This will provide a basis to further validate pre-identified characteristics and motivations on a new dataset, whilst simultaneously providing the opportunity to identify new concepts that may have not previously been investigated.





\bibliographystyle{model1-num-names}
\newpage\pagestyle{plain}
\bibliography{review.bib}


\end{document}